\documentstyle[general_cite,psfig]{mn} 
\newcommand{\lapprox}{\lower0.8ex\hbox{$\buildrel <\over\sim$}}
\newcommand{\gapprox}{\lower0.8ex\hbox{$\buildrel >\over\sim$}}

\title[The Protoelliptical NGC 1700]{Imaging of the protoelliptical
  NGC~1700 and its globular cluster system}

\author[R. J. N. Brown et al.]
{Richard J. N. Brown,$^1$ Duncan A. Forbes,$^{1,2}$ Markus Kissler--Patig$^3$\cr and Jean P. Brodie$^4$\\ 
$^1$ School of Physics and Astronomy, University of Birmingham,
Edgbaston, Birmingham, B15 2TT, UK\\
Email: rjnb, forbes@star.sr.bham.ac.uk\\
$^2$ Astrophysics \& Supercomputing, Swinburne University of Technology, Hawthorn NSW 3122, Australia\\
dforbes@swin.edu.au\\
$^3$ ESO, Karl--Schwarzchild--Strasse 2, D--85748, Garching bei M\"{u}nchen, Germany\\
mkissler@eso.org\\
$^4$ Lick Observatory, University of California, Santa Cruz, CA 95064, USA\\
brodie@ucolick.org\\}
\date{Accepted .....................; Received .....................;
in original form .......................}
\pagerange{\pageref{firstpage}--\pageref{lastpage}}
\pubyear{2000}

\begin{document}

\maketitle
\label{firstpage}
\begin{abstract}
  
  An excellent candidate for a young elliptical, or `protoelliptical'
  galaxy is NGC~1700. Here we present new $B$, $V$ and $I$ band imaging
  using the Keck telescope and reanalyse existing $V$ and $I$ band images
  from the \emph{Hubble Space Telescope}. After subtracting a model of the
  galaxy from the Keck images NGC~1700 reveals two symmetric tidal
  tail-like structures. If this interpretation is correct, it suggests a
  past merger event involving two spiral galaxies.  These tails are largely
  responsible for the `boxiness' of the galaxy isophotes observed at a
  radius of $\sim13$ kpc.
  
  We also show that the $B-I$ colour distribution of the globular cluster
  system is bimodal.  The mean colour of the blue population is consistent
  with those of old Galactic globular clusters.  Relative to this old,
  metal poor population, we find that the red population is younger and
  more metal rich.  This young population has a similar age and metallicity
  as that inferred for the central stars, suggesting that they are both
  associated with an episode of star formation triggered by the merger
  that may have formed the galaxy.  Although possessing large errors, we
  find that the majority of the age estimates of NGC~1700 are reasonably
  consistent and we adopt a `best estimate' for the age of $3.0\pm1.0$ Gyr.
  This relatively young age places NGC~1700 within the age range where
  there is a notable lack of obvious candidates for protoellipticals.  The
  total globular cluster specific frequency is rather low for a typical
  elliptical, even after taking into account fading of the galaxy over the
  next 10 Gyr.  We speculate that NGC~1700 will eventually form a
  relatively `globular cluster poor' elliptical galaxy.

\end{abstract}


%

\section{Introduction}

Merging is now thought to be a key process in the evolution of galaxies.
The hypothesis that two colliding spiral galaxies will eventually form an
elliptical galaxy \cite{toomre72} has gained much observational and
theoretical support over the years. The `smoking gun' of this type of
merger is the presence of two tidal tails, formed from the progenitor's
discs. A number of well known examples are found in the local Universe,
such as {\it The Mice}, {\it The Antennae}, Arp~220, NGC~3256 and NGC~7252.
For such galaxies a variety of methods are available (e.g. spectroscopy of
the stellar populations, dynamical measurements, comparison with models
etc.) to estimate the time since the merger occurred. The derived ages, for
these classic tidal tail systems, is up to 1--2 Gyr since nuclear
coalescence. It was Ivan King in 1977 who first pointed out the general
lack of obvious candidates for older merger remnants (i.e. 2--5 Gyr old).
These galaxies have been referred to as `King gap objects' or
protoellipticals. Identifying, and age dating, these protoellipticals could
provide the missing link between late stage spiral mergers and elliptical
galaxies.

Indeed, a crucial step in testing the merger hypothesis would be to show an
`evolutionary consistency', i.e. that spiral mergers and protoellipticals
evolve to have the same energetic, structural, dynamical and chemical
properties as normal, old ellipticals. This has been problematic due to the
difficulty of estimating the age of old stellar populations, without
telltale morphological signatures such as tidal tails.  Recently, several
different methods have become available. These include breaking the
age--metallicity degeneracy with spectral synthesis (e.g.
\pcite{gonzalez92}) and new models (e.g. \pcite{wortheymod}), quantifying
optical fine structure \cite{ss92} and using the colours of globular
clusters (e.g. \pcite{whitmore97}).

An ideal candidate for a nearby protoelliptical is NGC~1700. It
possesses a kinematically distinct core (Franx, Illingworth \& Heckman
1998a), reveals evidence for extensive morphological disturbance
\cite{ft92} and it has a high rotational velocity to velocity
dispersion ratio, relative to other ellipticals (Bender, Burstein \&
Faber 1992).  As well as being a possible protoelliptical, it offers
the opportunity to compare various age estimates. For example, age
estimates can be made from its two faint tidal tails, optical fine
structure, globular cluster colours, kinematic structure, and its
stellar component.

We have adopted the same distance to NGC~1700 as \scite{whitmore97},
i.e.  51.4 Mpc (which includes a correction for Virgocentric infall
and assumes $H_{\circ}=75$ km~s$^{-1}$~Mpc$^{-1}$). This corresponds
to 249 pc per arcsec. The total $B$ band magnitude for NGC~1700 is
$M_{\mathrm{B}}=-21.56$ (RC3:\pcite{rc3}). It is classified as an E4
elliptical in the RC3 and E3 in the RSA. Photometric studies have been
carried out by Franx, Illingworth \& Heckman (1998b) and
\scite{g94}. A recent kinematic study is that of Bender, Saglia \&
Gerhard (1994).

In this paper we reanalyse {\it Hubble Space Telescope} (\emph{HST}) WFPC2 $V$, $I$
images and present new Keck $B$, $V$ and $I$ images. We focus on the morphology
of the galaxy and photometry of the globular cluster system, both of
which provide new age estimates. We compare our age estimates with a
variety of alternative methods and discuss the implications of our results
for the formation and evolution of elliptical galaxies.

\section{Observations and initial data reduction}

\subsection{Keck data}

New $B$, $V$ and $I$ images of NGC~1700 were obtained using the 10 m
Keck--I telescope at the W. M. Keck Observatory, Mauna Kea, Hawaii.  The
observations were carried out on 1997 September~30th, with seeing in each
filter of $\sim1.3$~arcsec. The LRIS instrument was used with the Tek
$2048\times2048$ CCD, giving a rectangular field--of--view of $6\times 8$
arcmin and a pixel scale of 0.215~arcsec~\mbox{pixel$^{-1}$}.  A single
long (600~s) exposure was taken in $V$ and $I$ filters and two 600~s
exposures were taken in the $B$ band. We also took a short (10~s) exposure
in each filter. For the purpose of photometric calibration we obtained
exposures in each filter of the field SA98 from \scite{landolt92}.

The data were reduced in a standard manner using \textsc{iraf} software.
After bias subtraction there remained a small offset between the two halves
of the CCD.  This is due to the dual amplifier mode of the CCD read-out.
This effect is reasonably well approximated by a step function which varies
perpendicular to the read-out direction. The bulk of this offset was
successfully removed after division by the normalised flat-field.  A small
offset ($\sim1-3$ per cent) remained, which was corrected for by
multiplying one side by a factor determined by taking the mean values of
several sky regions either side of the offset. This successfully removed
the offset for most of the image. However, an offset remained in the bright
regions at the very centre of the galaxy ($\sim$~few arcsec). This may be
due to some non-linearity in the gains of the amplifiers at high count
rates (as suggested in the LRIS manual). This problem clearly affects
galaxy photometry in the inner regions but as described below, does not
affect standard star or globular cluster (GC) photometry.

Photometric calibration was performed using our standard star exposures of
SA98 and the photometric measurements of \scite{landolt92}. The resulting
rms error in the final photometry is $\pm0.02$ mag in all three bands.
K-correction and Galactic extinction were included using the same values
for NGC~1700 as in \scite{g94}, 0.18, 0.11 and 0.07 in the $B$, $V$ and $I$
bands respectively. In order to investigate the random errors on our
zeropoints we compared the measurements for the same stars that appear in
each pointing. We found a random error of $0.02$ mag. To better quantify the
systematic error due to the dual amplifier offset, we compared the mean
zeropoints for the stars on the left and stars on the right of the offset
in the same frame. We found a systematic error due to the offset of
$0.03$ mag.

\section{Data analysis}
\subsection{\emph{HST} data}
The $V$ and $I$ band \emph{HST} data of NGC~1700 were obtained from proposal
\mbox{G0--5416} in the \emph{HST} archive.  Photometry of the GCs in NGC~1700 has
been discussed by \scite{whitmore97}. Here we reanalyse the \emph{HST} data, and
use the results to aid with GC selection in our Keck data.

We performed the photometry in 2~pixel radius apertures with
\textsc{sextractor}
\cite{bertin96}, applying the aperture correction given by
\scite{whitmore97} and the calibration and transformation to Johnson
magnitudes following \scite{holtzman95}.

From the initial object list of 383 sources detected by \textsc{sextractor} we
applied the following selection criteria: The magnitude
range was restricted to be $21.5<V<26.5$, the faint limit chosen to avoid
introducing colour bias effects due to lack of completeness. Galactic GCs
at the distance of NGC~1700 are expected to be at magnitudes of $V>22.5$.
However, in order to account for the possible presence of a younger
population of GCs that might be up to $\sim1$ mag brighter than those in
the Milky Way, we relaxed our bright limit to be $V=21.5$.

The object colours were restricted to $0.6<V-I<1.7$ (which is roughly
equivalent to $-2.5< [{\mathit{Fe}}/H] <+1$). Our typical error is $\pm0.1$ in
magnitude and $\pm0.15$ in colour. Finally we carried out a careful
visual check of the objects on the image display in order to remove
point-like or galaxy-like objects. The final list contains a total of
146 GCs from all 4 WFPC2 chips.

\subsection{Keck data}
\subsubsection{Galaxy morphology}

In order to better reveal the fine structure present within NGC~1700 it is
necessary to subtract off a model of the underlying galaxy. All modelling
was performed using the \textsc{isophote} package in \textsc{stsdas} (e.g.
\pcite{ft92}). We initially constructed a model for the purpose of sky
subtraction.  Elliptical isophotes were fitted with a fixed centre and
employing a \mbox{3-$\sigma$} clipping algorithm. The ellipticity and position
angle of the isophotes were allowed to vary freely. A pixel mask file was
also constructed to explicitly mask out stars and pixel defects.  We were
unable to perform fits for the long exposure images at small radii because
the central regions were saturated.  For these images we set the initial
fit radius beyond the saturated region.  The fits were allowed to progress
outwards until constraints on signal to noise $(S/N)$ terminated the
process.  Sky subtraction was performed in a similar way to that detailed
in \scite{g94}, i.e.  the model intensity profiles in the outer parts of
the galaxy were fitted by a power-law in order to accurately determine the
sky level.  The appropriate level was then subtracted from each image. The
sky-subtracted images were then re-modelled in an identical way to before,
to reveal the structure within the body of the galaxy. The models typically
reached out to a distance of $\sim180$~arcsec (45~kpc) from the galaxy
centre.

Our fits extend significantly further from the centre than the imaging
of either \scite{g94} or \scite{franx89b}. Using the short exposure
image, we estimate a total $B$ band magnitude for the galaxy of
$12.00\pm0.05$, in excellent agreement with that given by the RC3 of
12.01. We also estimate a total $V$ magnitude of $11.18\pm0.05$ which
agrees reasonably well with the RC3 value of 11.10, and a total $I$
magnitude of $9.80\pm0.10$ compared with 9.87 from \scite{poulain88}.

\begin{figure}
\begin{minipage}{8.7cm}
\centering
\psfig{file=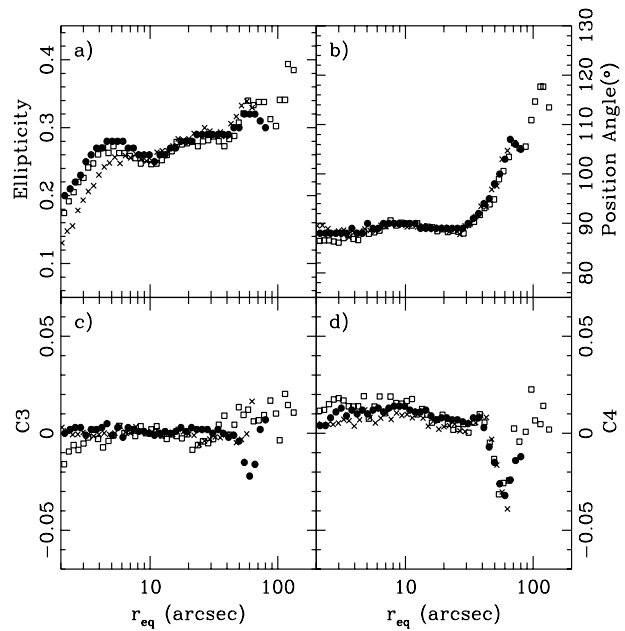,angle=0,height=8.7cm,width=8.7cm}
\end{minipage}
\caption{Radial profiles of the fit to our short $B$ band exposure (open squares) in terms of equivalent radius ($r_{\mathrm{eq}}=a\sqrt{1-\epsilon}$). For comparison, the profiles of \protect\scite{g94} and \protect\scite{franx89b} are shown as crosses and solid circles respectively. Our results agree well with these previous studies but extend to larger radii.}
\label{fig:comp}
\end{figure}

We show in Fig.~\ref{fig:comp} the radial profiles of the ellipticity,
position angle (PA), third cosine term (C3) and the fourth cosine term (C4)
of the fitted isophotes for the short $B$ band image. All radii are
expressed in terms of `equivalent radius',
$r_{\mathrm{eq}}=a\sqrt{1-\epsilon}$, where $a$ is the major axis radius
and $\epsilon$ is the isophote ellipticity. The C4 term is perhaps the most
interesting of a Fourier series which expresses the deviations from a pure
ellipse at a given radius. This term describes whether the isophotes are
discy (C4 is positive) with excess light along the major and/or minor axes,
or boxy (C4 is negative), representing excess light at 45\degr\, with
respect to these axes. There is good agreement between the shape and
absolute values of our data and that of \scite{g94} and \scite{franx89b}.
The slight discrepancies at small radii are almost certainly due to seeing
effects. Our profiles in $V$ and $I$ were similar in both shape and
absolute value to that in the $B$ band.

\subsubsection{Globular cluster sample selection and photometry}
As with the \emph{HST} data, potential GC candidates were detected in the
Keck images using the \textsc{sextractor} program. \textsc{sextractor}
provides measures of FWHM, ellipticity and an indication of whether an
object was `pointy' (star-like) or extended (galaxy-like).  We found 615
candidates in common between the $B$, $V$ and $I$ band images.  Using this
object list, we performed photometry with the \textsc{iraf} task
\textsc{phot} . After a curve of growth analysis on several sources we
determined an optimum aperture size of 8 pixels with a suitable background
annulus of 15 to 20 pixels. In order to check the photometry and to
determine suitable selection criteria, we examined the 34 sources in common
between the $V$ band Keck image and our final $V$ band \emph{HST} globular
cluster list.  Similarly, 14 candidates were found in common between our
final \emph{HST} list and the Keck $I$ band image.  Although there was
large scatter between the \emph{HST} and Keck photometry at faint
magnitudes, there was no obvious systematic bias. Our selection criteria
are determined from the sample of 34 $V$ band sources in common between the
\emph{HST} and the Keck frames.  Due to \emph{HST}'s high spatial
resolution, we are confident that the vast majority of sources that
appear in the final \emph{HST} list are \emph{bona fide} GCs and hence
use them to refine our Keck candidate list.

\begin{figure}
\begin{minipage}{8.7cm}
\centering
\psfig{file=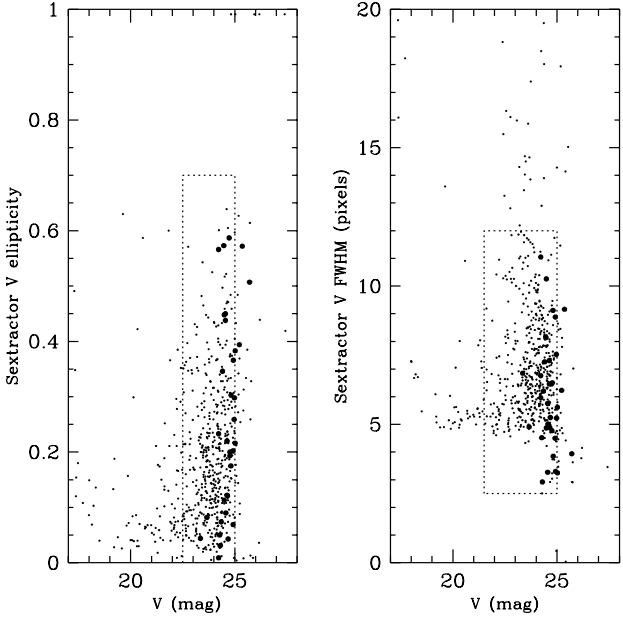,angle=0,height=8.7cm,width=8.7cm}
\end{minipage}
\caption{$V$ band ellipticities and FWHMs of the initial sample of 615 globular cluster candidates in common between the Keck $B$, $V$ and $I$ band images. The selection cuts applied to this sample are represented by the dotted boxes (see text for details). The large solid circles are the 34 globular clusters in common with the final \emph{HST} globular cluster list.}
\label{fig:cuts}
\end{figure}

The following criteria were applied to the 615 objects in common between
the Keck $B$, $V$ and $I$ frames: (a) $21.5<V<25.0$, (b)
$2.5<\mathrm{FWHM}<12.0$ (c) ellipticity $<0.7$.  These selection cuts are
shown in Fig.~\ref{fig:cuts}. As with our \emph{HST} magnitude selection,
the bright limit was chosen to include a population of young GCs while
excluding bright foreground stars. Our faint magnitude cut-off was chosen
in order to avoid introducing any colour bias into our results.  After
magnitude selection, 507 candidate GCs remained in our sample.

The selection using the \textsc{sextractor} FWHM and ellipticity measures
excluded a further 30 sources. These selection cuts were based on the shape
parameters observed for the GCs in common with the \emph{HST} list. We also
excluded a further 25 sources because their $x$ or $y$ centre positions as
determined by \textsc{phot} deviated from those originally found by
\textsc{sextractor} by more than 2 pixels, and were thus liable to be
misidentifications.

\begin{figure}
\begin{minipage}{8.7cm}
\centering
\psfig{file=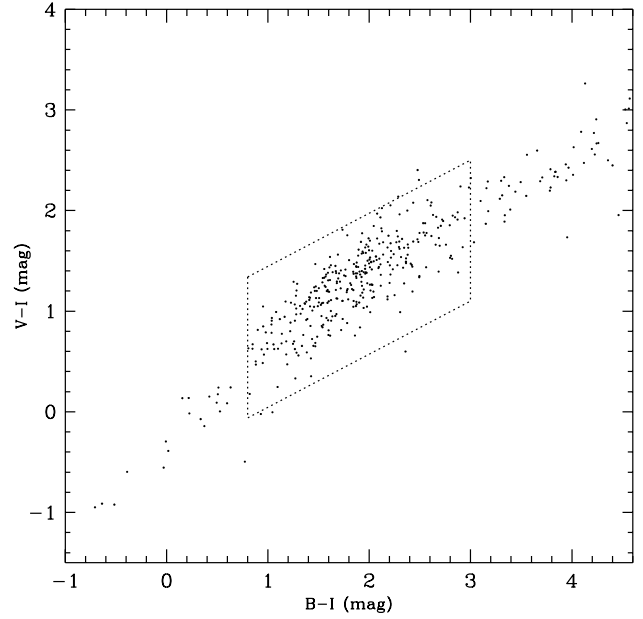,angle=0,height=8.7cm,width=8.7cm}
\end{minipage}
\caption{Colour--colour diagram for 452 globular clusters selected from our Keck data based on $V$ magnitude, FWHM, ellipticity and $x$, $y$ position selection (see text). The dotted box represents the selection based on colour and contains 352 globular clusters.}
\label{fig:colcut}
\end{figure}

The full range of expected GC metallicities is
$-2.5<[{\mathit{Fe}}/H]<+1.0$.  Using the Galactic colour--metallicity
relation of Couture, Harris \& Allwright (1990), this corresponds to a
range in colour of $1.2<B-I<2.5$. The errors in our photometry lead to
an average error in colour of $\sim\pm0.4$ mag. We thus decided to
relax our final colour selection to $0.8<B-I<3.0$. Our final colour
selection is shown by the dotted box in Fig.~\ref{fig:colcut}.  Here
the range in $V-I$ is defined as linear functions of $B-I$. This
colour selection reduced our sample size down to a list of 352
objects. As a final stage in selection we performed a visual check of
all the 352 candidate GCs in our colour selected list. From this
inspection we rejected a further 40 sources that resembled galaxies
(i.e. appeared diffuse and/or ellipsoidal) leaving a final list of 312
GCs.

Although these selection cuts are designed to reduce contamination to a
minimum, there is always the concern that our final GC sample will still be
contaminated by background galaxies and foreground stars. An estimate of
the number of contaminating sources can be made by taking an image of a
nearby `blank' field.  However, as we did not have such an image, we had to
use an alternative method. \scite{bahcall81} give the predicted stellar
densities in 17 fields based on their model of the Milky Way. Using the
star densities in the field closest in direction to our observations of
NGC~1700 (field 13), and correcting for our field--of--view, we predict
only 41 stars for our magnitude range of $21.5<V<25.0$. Similarly we
predict a total of 54 stars in our $I$ band field (using limiting
magnitudes of $20.5<I<23.5$). The equivalent stellar density for the $B$
band predicts only 29 stars in our field (assuming $B$ band limiting
magnitudes of $22.0<B<25.5$). As our selection criteria requires sources to
be in \emph{each} of our $B$, $V$ and $I$ images, we conclude that $\leq29$
foreground stars that could be present do not make up a significant
contribution to our candidate sample.

Differential galaxy counts in the $B$ band are given in fig.~1 of
\scite{kookron92}. From this we estimate a total of 1343 background
galaxies in our $B$ band frame down to our limiting magnitude of $B=25.5$.
Using \scite{yoshii93} we estimate a total of 1130 galaxies in our $I$ band
frame (to $I=23.5$). Once again, as we selected only those sources in
common between the $B$, $V$ and $I$ frames we can take the lower of these
estimates as an indication of the number of contaminating galaxies in our
list. These numbers are high compared to the 615 objects detected in common
between our $B$, $V$ and $I$ frames, though the number of background galaxies
actually detected will be significantly lower due to the incompleteness of
the sample at faint magnitudes. Although potentially large in number, we
show in Section~\ref{sec:GCcolour} that the additional constraint of colour
effectively excludes the majority of galaxies from our candidate GC list.

\section{Results and discussion}
\subsection{\emph{HST} data}
\label{sec:hstresults}

\begin{figure}
\begin{minipage}{8.4cm}
\centering 
\psfig{file=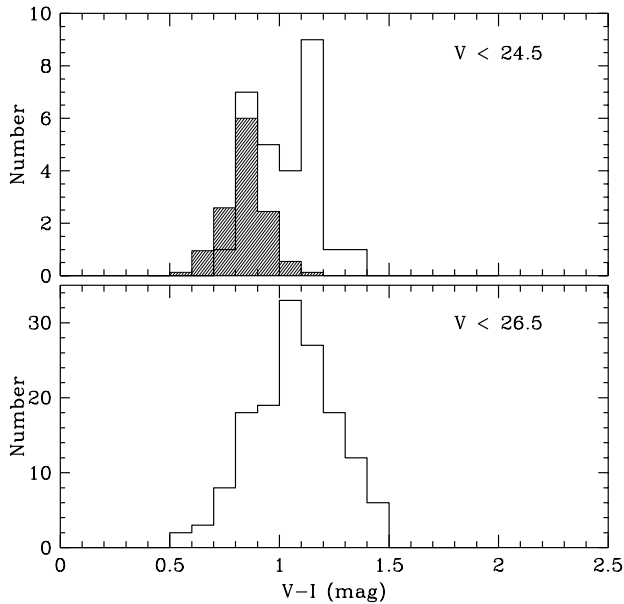,angle=0,height=8.4cm,width=8.4cm}
\end{minipage}
\caption{Histograms of NGC~1700 globular cluster $V-I$ colours from \emph{HST} images. The lower panel shows the $V-I$ distribution of our whole sample. The peak is at $V-I=1.07$. The upper panel shows the colour distribution for those globular clusters with $V<24.5$ (which have colour errors of $\sim\pm0.1$ mag) and shows evidence for two populations, one at $V-I=0.85\pm0.05$ and one at $V-I=1.15\pm0.05$. The scaled $(V-I)_0$ colour distribution for Galactic GCs is shown as a shaded region and is consistent with the blue peak in our bright sub-sample.}
\label{fig:hsthist}
\end{figure}

Two $V-I$ histograms for the GCs in our \emph{HST} sample are shown in
Fig.~\ref{fig:hsthist}. The lower panel shows the colour histogram for
the whole 146 GC sample. There is a single peak at around $V-I=1.07$,
which is consistent with that of \scite{whitmore97},
i.e. $V-I=1.05$. The upper panel shows the $V-I$ colour distribution
for GCs in our sample that are brighter than $V=24.5$. At $V=24.5$,
our typical error in colour is $\sim0.1$ mag. Any fainter than this
and the two distributions merge into a single broad peak due to
photometric errors. For the bright sample, there is a suggestion of
the presence of two peaks separated by $\Delta(V-I)=0.30\pm0.07$, with
one peak at $V-I=0.85\pm0.05$ and one at $V-I=1.15\pm0.05$. As we have
a small number of GCs brighter than $V=24.5$ we use the dip statistic
(see \pcite{hartigan85,gebhardt99}) in order to confirm this
bimodality. For our bright sample of \emph{HST} GCs, the dip statistic
indicates a probability of $>90$~per cent that the distribution is not
unimodal. The two peaks appear to have been `washed out' at fainter
magnitudes due to the photometric errors. The shaded region in this
figure represents the $(V-I)_0$ histogram of Galactic GCs. This
distribution shows a sharp peak at $(V-I)_0\sim0.9$, i.e.  slightly
bluer than the peak of the full sample but consistent with the blue
peak of the $V<24.5$ population.

An \emph{HST} study of the NGC~1700 GC system prior to that of \scite{whitmore97}
was performed by \scite{forbes96}. Although only 39 GCs were detected,
there was a hint of possible bimodality in the $V-I$ histogram with peaks
at $V-I\sim0.9$ and $V-I\sim1.2$. Another \emph{HST} observation of NGC~1700 was
carried out by Richstone et al. (see \pcite{gebhardt99}). An analysis of
these observations detected 27 GCs, again with two possible peaks in the
colour distribution at $V-I\sim0.9$ and $V-I\sim1.2$. However, a
statistical analysis of the $V-I$ distribution detected no significant
bimodality \cite{gebhardt99}.  The positions of the peaks in
Fig.~\ref{fig:hsthist} are consistent with those suggested by the previous
observations mentioned above.

\subsection{Keck data}
\subsubsection{Galaxy morphology and age estimates}
\label{sec:galmorph}
\begin{figure}
\begin{minipage}{8cm}
\centering 
\psfig{file=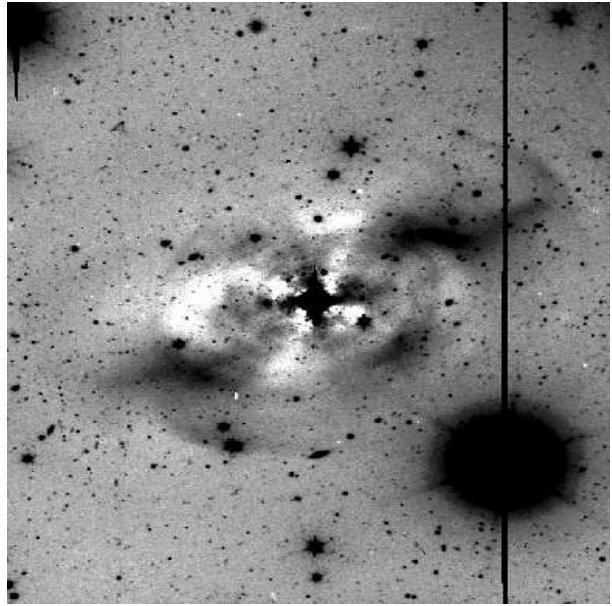,angle=0,height=8cm,width=8cm}
\end{minipage}
\caption{Residual map of NGC~1700 in the $V$ band after subtraction of a galaxy model. Dark features represent excess light. The
  image reveals two roughly symmetric tail-like structures to the
  North-West and South-East of the galaxy. Note also the saturated region
  in the centre.  The background has been raised to match the intensity at
  the outer modelling radius. The outermost modelled isophote is at a
 semi-major axis distance of 180~arcsec (45 kpc) from the centre.}
\label{fig:resid}
\end{figure}

In Fig.~\ref{fig:resid} we show the `residual image' of NGC~1700
produced by subtracting the model from the 600~s $V$ band image. Here
the fine structure within the galaxy is better revealed. Most notable
are the two broad, faint tidal tails or plumes visible to the
North-West and South-East of the galaxy. In addition many GCs are
apparent. The projected extent of the tidal features is about 165
arcsec (41 kpc) from the galaxy centre.  A faint shell system is just
visible in the central region (i.e.  within the central $\sim25$
arcsec) of the galaxy, as found in
\scite{ft92}.

\begin{figure}
\begin{minipage}{8.7cm}
\centering
\psfig{file=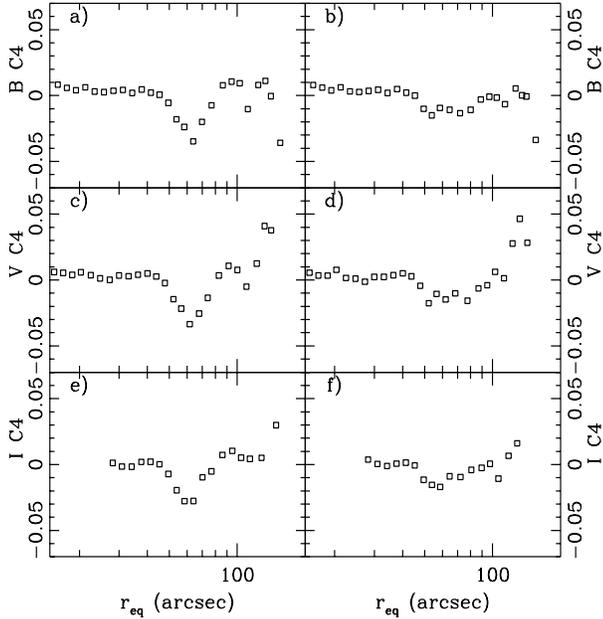,angle=0,height=8.7cm,width=8.7cm}
\end{minipage}
\caption{Radial profiles in the $B$, $V$ and $I$ bands of the 4th cosine term. a), c) and e) are profiles from the fits without masking the tails, while b), d) and f) are the profiles after masking. The boxiness at $\sim60$ arcsec is significantly reduced by excluding the tidal tails from the galaxy model.}
\label{fig:tailprofs}
\end{figure}

In Fig.~\ref{fig:tailprofs} we show the profile of the 4th cosine (C4)
parameter in the $B$, $V$ and $I$ bands for the long exposure images. The
C4 term is generally positive at radii $\lapprox30$ arcsec, representing
discy isophotes. At greater than 30 arcsec the isophotes become strongly
boxy, indicating an excess of light at $\sim45\degr$ from the major and
minor axes.  Fig.~\ref{fig:resid} shows that this is very probably due to
some light from the tail-like structures which remains in the galaxy
model. To investigate this further, the `tails' were masked out and the
galaxy remodelled.  Fig.~\ref{fig:tailprofs} also shows the resulting C4
profiles after masking the tails. Note that the boxiness in the original
fits has been greatly reduced in all cases (to less than $2$ per cent deviation),
indicating that the dominant cause of the boxiness at these radii is the
tails. The tails themselves appear clearly brighter in the residual maps
produced by subtraction of these models from the original images. This
shows that without masking, some of the light from the tails is included in
the model, modifying the fit parameters.

\scite{keelwu95} constructed a `merger sequence' from a sample of galaxies
that are good candidates for ongoing mergers and remnants of mergers
between two approximately equal mass disc galaxies. They assigned a
`merger stage' to each galaxy based upon dynamical crossing times,
with zero age defined to be the point of nuclear coalescence. They
noted that the fraction of galaxy starlight contained within the tails
roughly anti-correlates with this merger stage. This allows us make a
very rough estimate of the time-scale since the merger event that
created the tails within NGC~1700.

This analysis was performed on the long exposure residual images. As we
wanted to include as much of the tail light in the residual images as
possible, we used the residual images produced by excluding the tails from
the elliptical fit of the galaxy. As mentioned above, if the tails are not
masked out during the fit, some of their light is included in the model,
modifying the fit parameters and resulting in the subtraction of a
significant amount of tail light from the galaxy image. The total flux
contained within $\sim300$ small circular apertures positioned on the tails
was measured using the \textsc{iraf} utility \textsc{imexamine}. This was
used to derive a mean surface brightness for the tails, which was then
multiplied by their total area (i.e.  including the area missed due to
contaminating bright sources). The resulting surface brightnesses were
26.6, 25.9 and 24.4 mag~\mbox{arcsec$^{-2}$} for the $B$, $V$ and $I$ bands
respectively. The total light in the tails was divided by the total galaxy
light to give the `tail fraction'. We also roughly estimated the total tail
light using polygon-shaped apertures. This method gave a similar result to
the mean surface brightness method but included the light from
contaminating point sources and was thus less reliable.

\begin{figure}
\begin{minipage}{8.7cm}
\centering
\psfig{file=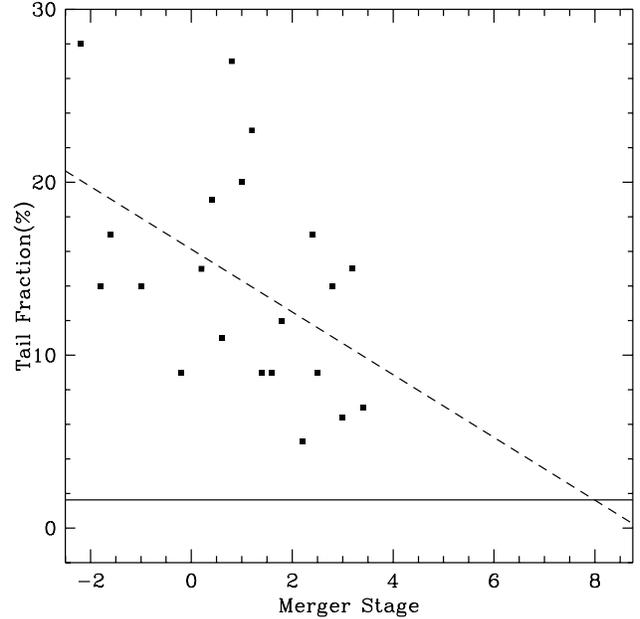,angle=0,height=8.7cm,width=8.7cm}
\end{minipage}
\caption{Tail fraction versus merger stage as derived from \protect\scite{keelwu95}. The dashed line represents a linear least-square fitted to the sample of galaxies in the merger sequence of Keel \& Wu (solid squares). The solid horizontal line represents the measured tail fraction of NGC~1700. From this method, we estimate that NGC~1700 is at a merger stage of $8.1\pm3.8$, which roughly corresponds to an age of $3.2\pm1.5$ Gyr.}
\label{fig:TF}
\end{figure}

The tail fractions in the $B$, $V$ and $I$ bands were found to be $1.84$,
$1.64$ and $1.72$ per cent respectively. The tail fractions in
\scite{keelwu95} were derived in the $V$ band, with a few exceptions. We
thus use our $V$ band tail fraction (which is similar to the $B$ and $I$
values) for NGC~1700 in the following analysis.  We fitted a linear
least-square to the points obtained from the \scite{keelwu95} data.  This
fit is shown as a dashed line in Fig.~\ref{fig:TF}. The solid horizontal
line represents the measured $V$ band tail fraction for NGC~1700. If we
extrapolate the fit we find that the $V$ band tail fraction measured for
NGC~1700 corresponds to a merger stage of $8.1\pm3.8$.  The large error on
this stage estimate is due to the scatter of the data points of
\scite{keelwu95}. We also performed a quadratic fit on the Keel~\& Wu data.
Extrapolating this fit yielded a stage comparable with the result using a
linear least-square fit.

By comparing the spectroscopic (i.e. central starburst) ages of several
galaxies in the \scite{keelwu95} sample with their assigned merger stage,
we determined the \emph{approximate} time since nuclear coalescence.  For
the stage of NGC~1700, the time since the central starburst and hence tail
formation is approximately $3.2\pm1.5$ Gyr.

A lower limit for the age of the tidal tails can be estimated from the
dynamical time-scale, i.e. $t_{\mathrm{dyn}}\sim(R/V)$. The projected distance
of the tidal tails from the galaxy centre is about 165 arcsec (41
kpc), and the rotation velocity in the outer parts is $\sim50$
km~\mbox{s$^{-1}$} \cite{bender94}. This gives a dynamical time of
0.8 Gyr, which is an underestimate if the tails do not lie in the
plane of the sky; thus the tidal tails are $>0.8$ Gyr old. This lower
limit is consistent with the age derived from the fraction of galaxy
light contained within the tails.

The presence of two symmetric tidal tails is generally taken to be a
signature of a recent major merger involving two, approximately equal mass
spiral galaxies (see e.g. \pcite{toomre72}). If the tidal features in
Fig.~\ref{fig:resid} are indeed genuine tidal tails then we could conclude
that NGC~1700 has experienced a major merger during its recent history.
Alternatively, if the features are merely plumes of tidally disturbed
material, the situation is less clear.  While a major merger could not be
ruled out, the situation of a disc galaxy merging into an existing
elliptical would be possible.

\subsubsection{Age estimates from globular cluster colours and magnitudes}
\label{sec:GCcolour}

\begin{figure}
\begin{minipage}{8.7cm}
\centering
\psfig{file=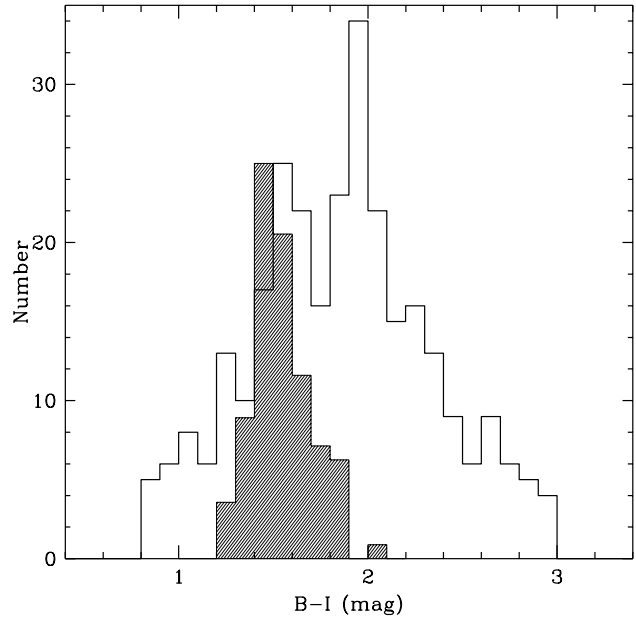,angle=0,height=8.7cm,width=8.7cm}
\end{minipage}
\caption{Histogram of $B-I$ colours for our sample of 312 globular clusters in NGC~1700 from Keck images. Two peaks are apparent with centres at $B-I=1.54\pm0.05$ and $B-I=1.98\pm0.05$ respectively. The shaded area represents the $(B-I)_0$ histogram of globular clusters in the Milky Way.}
\label{fig:BIhist}
\end{figure}

We present in Fig.~\ref{fig:BIhist} the $B-I$ colour histogram for
NGC~1700 GCs from the Keck data. The histogram appears bimodal with a
blue peak at $B-I=1.54\pm0.05$ and a second peak $0.44\pm0.07$
magnitudes redder at $B-I=1.98\pm0.05$. This bimodality does not
appear to be an artifact of the data binning and is still present if
the histogram bin boundaries are changed. For the subsequent
discussion and analysis we define the \emph{blue} population as those
GCs possessing $0.8<B-I\leq1.75$ and the \emph{red} population with
colours $1.75<B-I<3.0$. A statistical analysis using the \textsc{kmm}
algorithm (Ashman, Bird \& Zepf 1994) detects bimodality in the
distribution with $>99$ per cent confidence. The \textsc{kmm}
algorithm assigned a colour cut between the blue and red populations
of $B-I=1.8$, thus confirming our initial visual estimate. The $B-I$
distribution of GCs in the Milky Way is also shown in
Fig.~\ref{fig:BIhist} (shaded area). The peak of this distribution is
at $(B-I)_0\sim1.5$ which is similar to the peak of the blue
population of NGC~1700.

Is the bimodality of the $B-I$ histogram evidence for two distinct GC
populations in NGC~1700 ? In order to address this question, one has to
consider the expected number of contaminating sources within our final
sample. As the number of predicted stars \emph{before} colour selection is
small, we can immediately conclude that the contamination by foreground
stars in our final sample is negligible. Another source of contamination is
background galaxies. Our automatic and visual checks have removed obvious
galaxies but it is possible that small, unresolved background galaxies
remain. For our mean $B$ magnitude of 24.5, we expect to be detecting
sources out to a mean redshift of $z\sim0.8$ \cite{kookron92}.  At this
redshift all morphological types of galaxies (with the exception of
irregular types) have typical $B-I$ colours in excess of 2.5. This is
significantly redder than the peak of the red population and we can thus be
fairly confident that the two peaks are real and due to two distinct
populations of GCs. To support this view we performed photometry on 13
galaxy-like objects that appeared in our initial list but were rejected
during the selection process. The mean colour of these objects was
$B-I=2.7\pm0.2$; significantly redder than the peak of our red population
of objects in our final sample. Moreover, the mean FWHM value measured for
these objects was $12.3\pm1.2$~pixels; greater than the upper cut-off used
in our selection based on FWHM.

\label{sec:discussgcage}

The study of globular cluster systems can reveal important information
regarding the formation and evolution of the parent galaxy. The merger
model of galaxy formation (\pcite{ashman92}; \pcite{zepf93}) makes a number
of predictions about the properties of the GC system of the resultant
galaxy.  In this scenario, an elliptical galaxy is formed by the gas rich
merger of two spiral galaxies. If our interpretation of two, roughly
symmetric tidal tails in NGC~1700 is correct, then the galaxy has probably
undergone such an event. If this is the case, the merger model predicts the
formation of a new, metal rich population of GCs during the merger, which
then reddens and fades with time. Post-merger ellipticals would thus be
expected to possess a `new' metal rich population and an old, metal poor
population originating in the progenitor disc galaxies.
Fig.~\ref{fig:BIhist} reveals evidence for two GC populations in NGC~1700.
We next use the colours and magnitudes of these two populations to
determine their age and metallicity properties.

\begin{figure}
\begin{minipage}{8.7cm}
\centering
\psfig{file=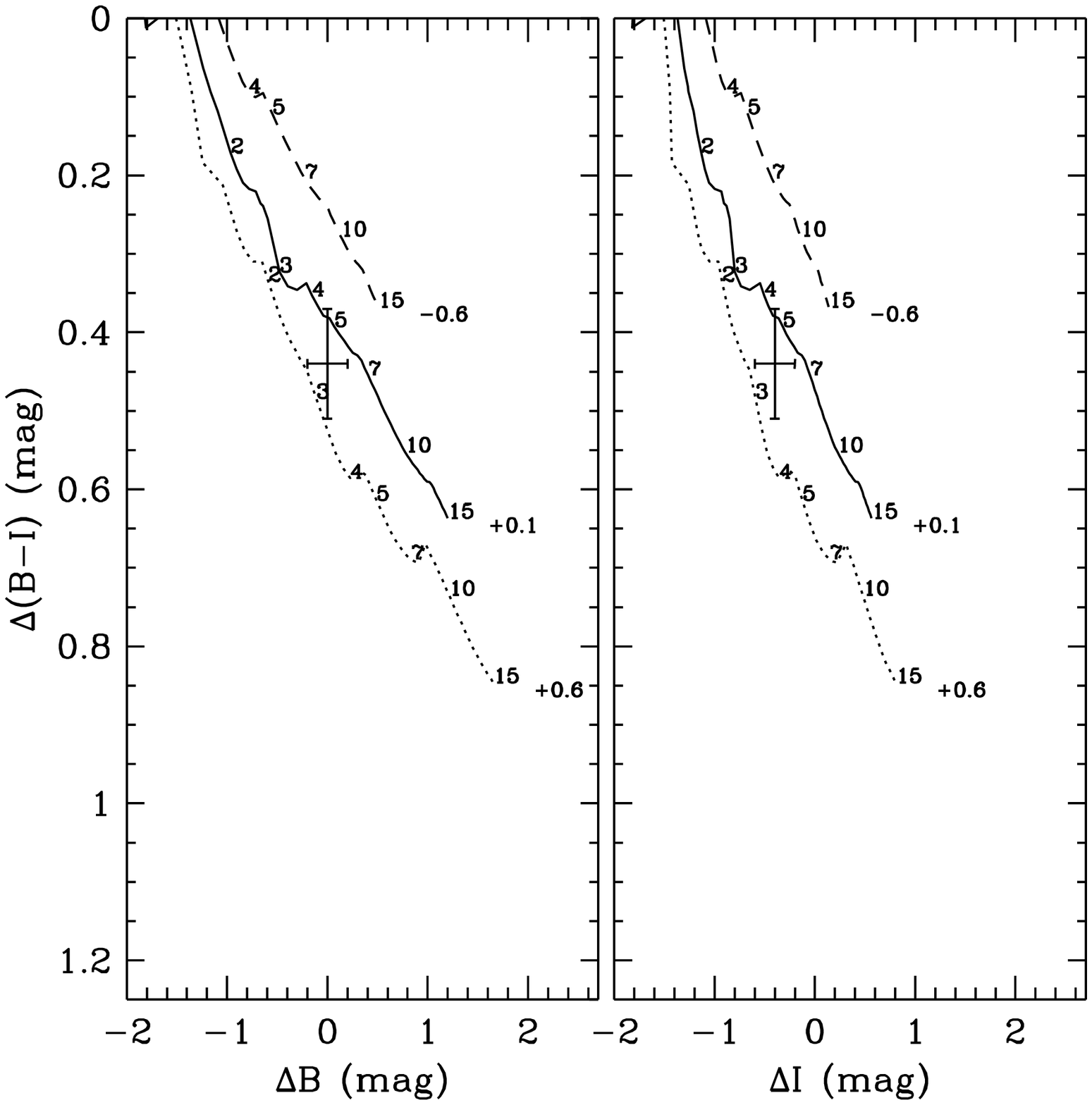,angle=0,height=8.7cm,width=8.7cm}
\end{minipage}
\caption{Evolution of $\Delta(B-I)$ vs. $\Delta B$ and $\Delta I$ with age from the models of \protect\scite{bc96}. The $\Delta(B-I)$ is defined as the colour difference between the red globular cluster population and a blue old, metal poor population with an age of 15 Gyr and $[{\mathit{Fe}}/H]=-1.5$. $\Delta B$ and $\Delta I$ are defined as the average magnitude difference between the two populations in the $B$ and $I$ bands respectively. The three tracks represent the evolution of the red population with metallicities as indicated on the end of each line. The cross shows the observed colour and magnitude difference between the two populations detected in NGC~1700, i.e. $\Delta(B-I)=0.44\pm0.07$, $\Delta B=0.0\pm0.2$ and $\Delta I=-0.4\pm0.2$. The predicted age of the red population is 2.5--5.0 Gyr with super-solar metallicity.}
\label{fig:dbididbbc96}
\end{figure}

\begin{figure}
\begin{minipage}{8.7cm}
\centering
\psfig{file=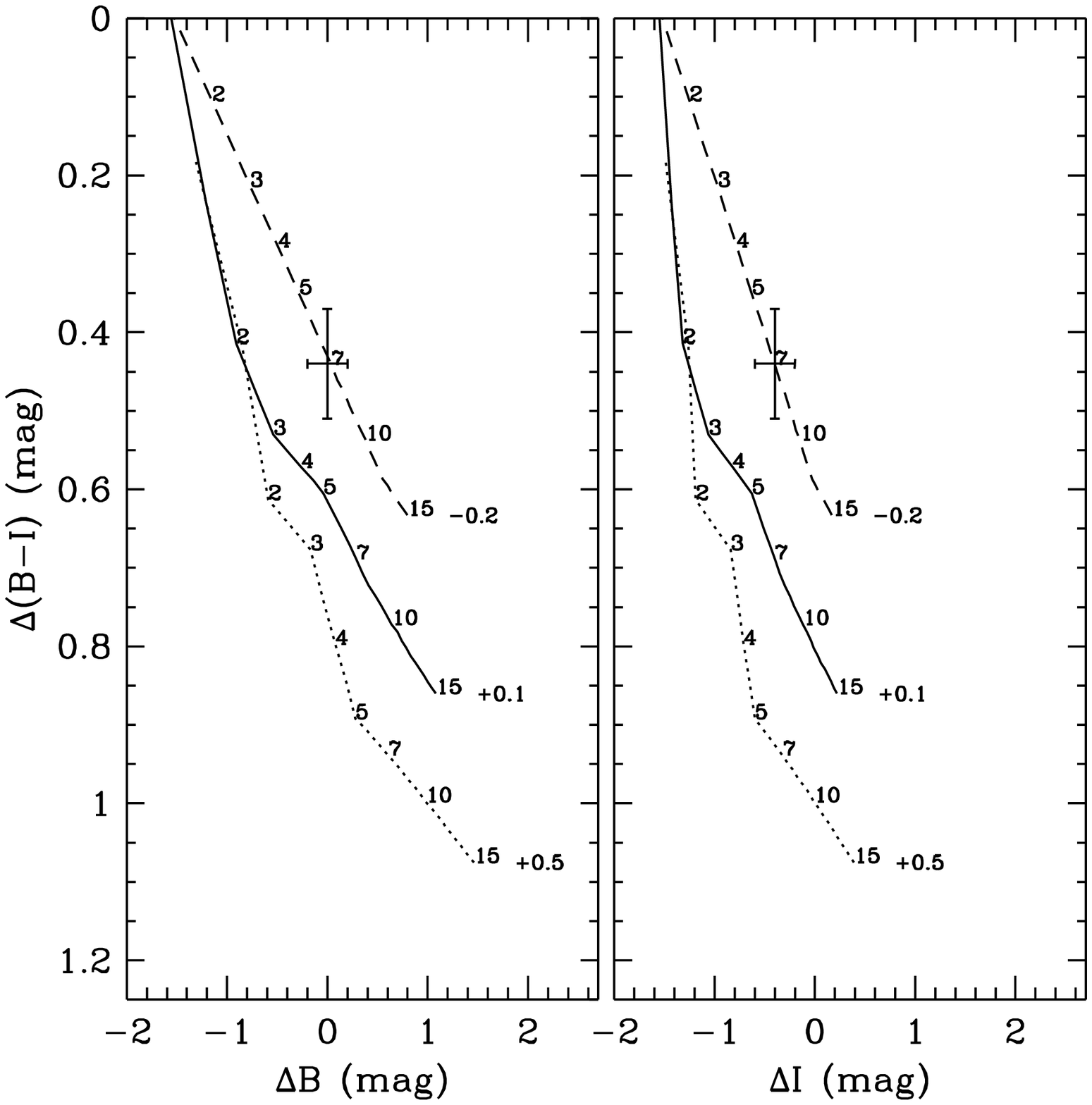,angle=0,height=8.7cm,width=8.7cm}
\end{minipage}
\caption{Evolution of $\Delta(B-I)$ vs. $\Delta B$ and $\Delta I$ with age
  from the models of \protect\scite{wortheymod}. The $\Delta(B-I)$ is defined as
  the colour difference between the red globular cluster population and a
  blue old, metal poor population with an age of 15 Gyr and
  $[{\mathit{Fe}/H}]=-1.5$.  $\Delta B$ and $\Delta I$ are defined as the
  average magnitude difference between the two populations in the $B$ and
  $I$ bands respectively. The three tracks represent the evolution of the
  red population with metallicities as indicated on the end of each line.
  The cross shows the observed colour and magnitude difference between the
  two populations detected in NGC~1700, i.e. $\Delta(B-I)=0.44\pm0.07$,
  $\Delta B=0.0\pm0.2$ and $\Delta I=-0.4\pm0.2$. The predicted age of the
  red population is 5.0--8.0 Gyr with a sub-solar metallicity.}
\label{fig:dbididbw94}
\end{figure}

The blue peak of our $B-I$ histogram is roughly consistent with the peak of
the $B-I$ distribution of Galactic GCs (see Fig.~\ref{fig:BIhist}). The
Milky Way is likely to be a typical example of a progenitor in a present
day gas-rich merger, thus indicating that we may assume that our blue peak
in $B-I$ is due to an old, metal poor population of GCs. Here we use the
stellar population models of both \scite{wortheymod} and \scite{bc96} to
predict the difference in magnitude and colour for a population of young
clusters, relative to an old, metal poor population with an age of 15 Gyr
and $[{\mathit{Fe}}/H]=-1.5$.  Fig.~\ref{fig:dbididbbc96} shows the predicted
differences in $\Delta(B-I)$ vs.  $\Delta B$ and $\Delta I$ based on the
models of \scite{bc96}.  Fig.~\ref{fig:dbididbw94} shows the equivalent
colour and magnitude changes based on the models of \scite{wortheymod}. In
each case, three different metallicity tracks are shown. From
Fig.~\ref{fig:BIhist}, $\Delta(B-I)=0.44\pm0.07$.

\begin{figure}
\begin{minipage}{8.4cm}
\centering
\psfig{file=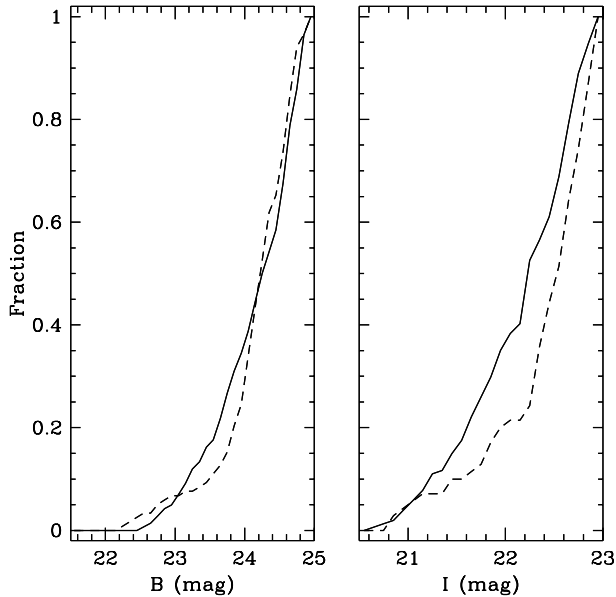,angle=0,height=8.4cm,width=8.4cm}
\end{minipage}
\caption{Normalised $B$ and $I$ band cumulative luminosity functions for the red (solid line) and blue (dashed line) populations of globular clusters. Only those globular clusters at least 0.5 magnitudes brighter than the corresponding limiting magnitudes are considered.}
\label{fig:cumulmag}
\end{figure}

The $\Delta B$ and $\Delta I$ values correspond to the magnitude
offset in the $B$ and $I$ bands between the young and old
populations. These offsets were estimated from the cumulative $B$ and
$I$ band GC luminosity functions for both the blue and red populations
separately. These luminosity functions are shown in
Fig.~\ref{fig:cumulmag}. To reduce the effect of photometric errors we
considered only those GCs at least 0.5 mag brighter than the limiting
magnitude in each band. The data has been normalised to the faint
limit of our luminosity functions. The median magnitudes of the red
and blue GCs considered were then calculated to allow us to determine
the separation of the blue and red populations.  The magnitude
differences between the red and blue populations were determined to be
$\Delta B=0.0\pm0.2$ and $\Delta I=-0.4\pm0.2$ in the $B$ and $I$
bands respectively. The observed values with their associated errors
are indicated by a cross on Fig.~\ref{fig:dbididbbc96} and
Fig.~\ref{fig:dbididbw94}.  Reference to the Bruzual \& Charlot models
shows that the red population is consistent with a young, metal rich
population of globular clusters, i.e. with an age of 2.5--5.0 Gyr and
super-solar metallicity ($[{\mathit{Fe}}/H]\sim+0.1$ to +0.6). A plot
of $\Delta V$ vs.  $\Delta(B-I)$, where $\Delta V=0.1\pm0.2$ predicts
a similar age and metallicity. It is interesting to note that the
Worthey models predict significantly different values for the age and
metallicity of the red GC population in the sense that they are older
and more metal poor. From the Worthey models the age of the red GCs is
5.0--8.0 Gyr and $[{\mathit{Fe}}/H]\sim-0.2$.  It thus appears that
ages derived from GC colours and magnitudes are highly model
dependent.  However both models suggest that the red GCs are both
younger and more metal rich than the blue population.  We may suspect
that $[{\mathit{Fe}}/H]$ should be at least solar metallicity. In the
merger model, the young population is thought to have formed from the
relatively enriched gas in the spiral discs, in contrast to the old
population which formed from metal poor gas.  Moreover, the line
strengths plotted in Fig.~\ref{fig:grids} (see
Section~\ref{sec:discussages} for details) imply for both models that
the stellar metallicity of the last major starburst is
$[{\mathit{Fe}}/H]\geq+0.5$. If the young GCs formed from the same
gas, we would expect them to be metal rich as well, which in turn
would favour the younger age of the Bruzual \& Charlot
models. However, if the young GCs were formed at an early stage of a
merger-induced starburst then they could in principle be more metal
poor than the present young stellar population
\cite{fritze95}. To summarise, if we assume that the blue GCs are an old
metal poor population, the red GCs are consistent with an age of
2.5--5.0 Gyr, assuming they have a metallicity of
$0.0\leq[{\mathit{Fe}}/H]\leq+0.5$.  In the case that the red GCs are
relatively metal poor, as suggested by the Worthey model tracks in
Fig.~\ref{fig:dbididbw94} then a slightly older (5.0--8.0 Gyr) age is
indicated. Infrared photometry or good spectra would help to resolve
the age and metallicity independently.

From Section~\ref{sec:hstresults} we found evidence for bimodality in $V-I$
with peaks at $V-I=0.85\pm0.05$ and $V-I=1.15\pm0.05$. A GC population with
an age of 15 Gyr, $[{\mathit{Fe}}/H]=-1.5$ and $V-I=0.85$ will have
$B-I\sim1.5$.  This is consistent with the blue population seen in our
$B-I$ colour histogram.  Similarly, a population with an age of 3 Gyr,
$[{\mathit{Fe}}/H]=+0.5$ and $V-I=1.15$ is expected to have $B-I\sim1.9$,
i.e.  consistent with the red peak of our $B-I$ distribution.  It thus
appears that the colours of the red and blue peaks in our $B-I$ histogram
are consistent with the the peaks hinted at in the \emph{HST} $V-I$ colour
distribution. The bimodality seen in independent data sets confirms the
presence of two GC populations in NGC~1700.

\begin{figure}
\begin{minipage}{8.7cm}
\centering
\psfig{file=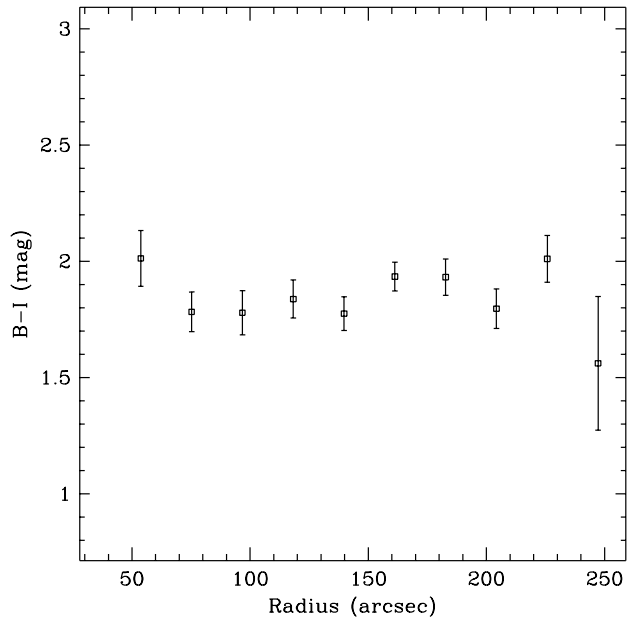,angle=0,height=8.7cm,width=8.7cm}
\end{minipage}
\caption{$B-I$ colour vs. galactocentric radius for globular clusters in NGC~1700. The data are binned into annuli of width 22 arcsec. Error bars correspond to Poisson statistics. Note the lack of any strong trend in colour with radius.}
\label{fig:colvrad}
\end{figure}

We show in Fig.~\ref{fig:colvrad} the mean $B-I$ colour of our GCs vs.
galactocentric radius. The data are radially binned into 10 annuli,
each of width 5.5 kpc (22 arcsec).  The GC system of NGC~1700 does not
show any significant correlation between colour and galactocentric
distance. In particular the red (i.e. young) GCs are not
preferentially concentrated towards the centre of the galaxy, as might
have been expected from the merger origin for new GCs
\cite{ashman92,zepf93}. A linear least-square fit to the radial colour
bins gives a gradient consistent with zero. We do not detect any
significant radial colour trend in the \emph{HST} sample of GCs
either, agreeing with the results of \scite{whitmore97}. This lack of
an obvious radial trend is potentially a problem for the merger origin
interpretation of the GCs, and deserves further consideration in any
future studies. It is also interesting to note that the GC systems of
several other ellipticals \emph{do} show radial colour gradients in
the sense that the red GCs are more centrally concentrated than the
blue (see e.g. Geisler, Lee \& Kim 1996; Forbes, Brodie \& Grillmair
1997).

\subsection{Other age estimates}
\label{sec:discussages}

Until recently, direct age dating of the stars in an elliptical was
near-impossible due to the well known degeneracy of age--metallicity
effects in old stellar populations. Spectral synthesis methods have
now been developed (e.g.  \pcite{wortheymod,bc96}) that show that
combinations of certain spectral line indices can efficiently
disentangle the effects of age and metallicity for young to
intermediate age stellar populations.  As well as stellar age dating,
a number of other methods are available for NGC~1700.  These estimates
are shown for comparison in Table~1. We show a mean spectroscopic age
based on the line strength data of Fisher, Franx \& Illingworth (1996) and
\scite{gonzalez92}, the age derived from the globular clusters by
\scite{whitmore97}, the `fine structure age' of \scite{ss92}, the age from
stellar dynamics (Statler, Smecker--Hane \& Cecil 1996) and the age
derived from the galaxy's deviation from the Fundamental Plane (see
Forbes, Ponman \& Brown 1998).

\begin{table*}
\begin{minipage}{14cm}
\caption{Summary of age estimates for NGC 1700}
\label{tab:ages}
\centering
\begin{tabular}{lcl}
\medskip Method & Age (Gyr)&Ref.\\
Globular clusters  & $4\pm2$ ? & Whitmore et al. (1997)\\
Fine structure  & $<6.0\pm2.3$ & Schweizer \& Seitzer (1992)\\
Stellar dynamics & $2.7-5.3$ & Statler et al. (1996)\\
Tail fraction & $3.2\pm1.5$ & This work\\
Globular clusters & $2.5-5.0$ & This work, Bruzual \& Charlot models (2000)\\
Globular clusters & $5.0-8.0$ & This work, Worthey (1994) models\\
Stellar spectroscopy & $2.9\pm0.3$ & This work, see text for details\\
Tail extent & $>0.8$ & This work\\
\medskip Fundamental Plane residual & $1.2\pm2.0$ & This work, Forbes et al. (1998b)\\
Best estimate & $3.0\pm1.0$ & This work\\
\end{tabular}

\end{minipage}
\end{table*}

\subsubsection{Ages from stellar population synthesis models}
\begin{figure}
\begin{minipage}{8.4cm}
 \centering \psfig{file=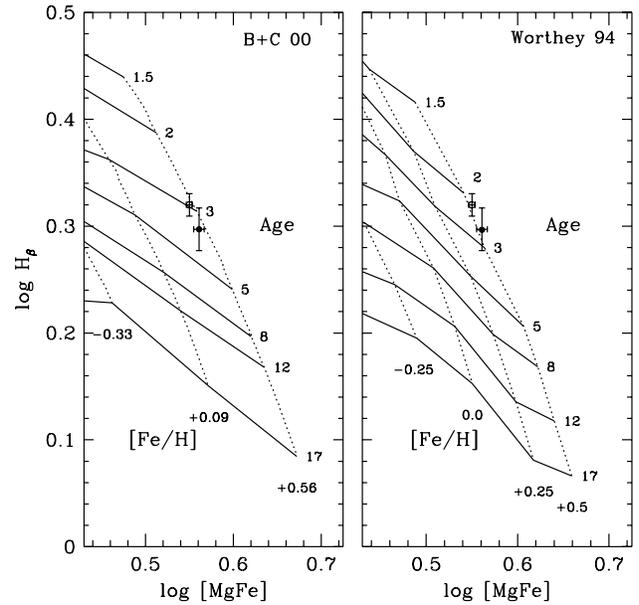,angle=0,height=8.4cm,width=8.4cm}
\end{minipage}
\caption{\protect\scite{wortheymod} and \protect\scite{bc96} model grids for 
$\log{\mathit{H\beta}}$ vs. $\log[{\mathit{MgFe}}]$.  The solid lines are
contours of constant age (in Gyrs) whereas the dashed lines represent lines
of constant metallicity. The solid and open circles indicate the positions
of NGC~1700 based on the line strength measurements of \protect\scite{fisher96} and
\protect\scite{gonzalez92} respectively.}
\label{fig:grids}
\end{figure}

The spectroscopic age estimates in Table~1 are based on comparisons of
observed absorption line strengths with single-burst stellar population
models of \scite{wortheymod} and \scite{bc96}. From these model grids,
observed line strengths can be used to give estimates of the age and
metallicity of the stellar component of a galaxy. We have used the
$\mathit{H\beta}$ and $[{\mathit{MgFe}}]$ line strengths from two sources,
i.e.  \scite{fisher96} and \scite{gonzalez92}. Fig.~\ref{fig:grids} shows
the $[{\mathit{MgFe}}]-{\mathit{H\beta}}$ model grids with the observed values
for NGC~1700 shown.  In both panels, the solid circle indicates the
observed line strengths from \scite{fisher96}, whereas the open circle
shows the line strengths from \scite{gonzalez92}.  The apertures used in
each case are $r_{\mathrm{e}}/10$ and $r_{\mathrm{e}}/8$ respectively
because we are primarily concerned with the age of the central starburst.
Using the Worthey models, the age derived from the Fisher et al. data is
$2.7\pm0.4$ Gyr and the age from the Gonzalez data is $2.3\pm0.3$ Gyr.  In
both cases, the metallicity, $[{\mathit{Fe}}/H]$ is $\geq+0.5$.  Similar
results are obtained by using the Bruzual \& Charlot models. In this case
the Fisher et al. data gives an estimated age of $3.5\pm0.5$ Gyr and the
data of Gonzalez suggests an age of $2.9\pm0.3$.  Combining these results
gives an average of $2.9\pm0.3$ Gyrs. Again both sets of data suggest a
metallicity of $[{\mathit{Fe}}/H]\geq+0.5$.  Note that this method measures a
`luminosity weighted mean age' of the central stellar population and is
thus likely to be dominated by the young stellar population associated with
the last episode of star formation. This may in turn be the result of a
merger induced starburst.

\subsubsection{Ages from globular clusters}

\scite{whitmore97} employed a method similar to that used in
Section~\ref{sec:discussgcage} to derive GC ages. This method relies on the
observed colour difference between two detected populations of GCs.
\scite{whitmore97} detected only one population (at $V-I=1.05$). Their age
of `$4\pm2$ Gyr ?' was derived from the non-detection of two distinct
populations, though it is consistent with the other age estimates discussed
herein. The non-detection appears to have arisen due to the small colour
separation of the GC populations. Using the same data but restricting the
sample to bright GCs, we have detected two separate populations as shown in
Fig.~\ref{fig:hsthist}. By including the whole sample (i.e. down to
$V=26.5$) the two peaks merge into a single distribution.

\subsubsection{Fine structure ages}
Another method of estimating galaxy age was developed by \scite{ss92}. They
defined a fine structure parameter, $\Sigma$, based on the amount of
optical `fine structure' present in a galaxy. This included a measure of
the maximum boxiness of the galaxy isophotes, the number and strength of
shells and tails and the presence or absence of `X-structure'. The
values of $\Sigma$ for the galaxies in their sample ranged from 0 for
galaxies with no fine structure to 7.6 indicating the largest amount of
fine structure observed. NGC~1700 was assigned a value of 3.70. They found
that $\Sigma$ correlated with a galaxy's residual from the mean
colour--magnitude relation. Using this fact and relating galaxy colours to
ages via a star formation model, they estimated the time since the merger
event. We quote their most representative age of 6.0 Gyr with an error of
$\pm2.3$ Gyr in Table~1. We note however , that the models used in
\scite{ss92} assumed a solar metallicity starburst. If the stellar
population is super-solar, as suggested by Fig.~\ref{fig:grids}, then the
\scite{ss92} age is expected to be a slight over-estimate.
  
\subsubsection{Dynamical considerations}
The age estimate of \scite{statler96} comes from measurements of the
stellar velocity field of NGC~1700. They found that within
$\sim2.5r_{\mathrm{e}}$ (8.8 kpc) the galaxy is kinematically well mixed.
This constrains the time since the last major merger event to be $\geq2.7$
Gyr to allow sufficient time for phase mixing and differential precession.
Their observations also define an upper limit for the time since the
merger.  The asymmetric photometric and kinematic signatures at larger
radii preclude a merger age greater than 5.3 Gyr, otherwise these features
would have relaxed and disappeared.  Thus we quote an age of 2.7--5.3 Gyr.
The same paper argues that the counter rotating core and boxy features (the
latter of which we attribute to the tidal tail-like structures) could not
have been created by the same merger event and that the observed form of
NGC~1700 must have arisen from the merger of at least three separate
stellar systems. However, it was not clear whether these events occurred
sequentially or simultaneously.  If the two tail-like structures are indeed
genuine tidal tails this would suggest that the history of NGC~1700 has
included at least one major merger event involving two approximately equal
mass disc galaxies.  If the kinematically distinct core (KDC) was formed
prior to this event it would have probably been disrupted during the
ensuing violent relaxation processes. If the KDC was indeed formed by a
separate process it must have resulted from a subsequent minor merger (e.g.
\pcite{balcells90}) or an interaction (see \pcite{thomson90}) some time
after the major merger event that created NGC~1700 and its tails.  Neither
of these suggestions are very appealing.  Alternatively, the tidal
structures seen in Fig.~\ref{fig:resid} could be interpreted as plumes
which have arisen from the infall of a small disc galaxy into a
pre-existing elliptical.

\subsubsection{Scatter from the Fundamental Plane}
A recent study by \scite{FP1} showed that a galaxy's deviation from the
Fundamental Plane (FP) correlated with its age, albeit with large scatter.
The FP residual is defined as
$R(\sigma_0,M_{\mathrm{B}},\mu_{\mathrm{e}})=2\log(\sigma_0)+0.286M_{\mathrm{B}}+0.2\mu_{\mathrm{e}}-3.101$
\cite{prugniel96}. Young ellipticals fall below the FP (i.e. have negative
residuals) and evolve towards it until they lie on the FP at an age of
about 10 Gyr. Older ellipticals tend to lie above the FP. It is thus
possible to use the FP residual to approximately age date an elliptical
galaxy. The FP residual for NGC~1700 is
$R(\sigma_0,M_{\mathrm{B}},\mu_{\mathrm{e}})=-0.37$. This corresponds to an
age of $1.2\pm2.0$ Gyr based on the fit in \scite{FP1}.  This is slightly
younger but comparable to the other age estimates discussed in this
section.

\subsubsection{Comparison of various age estimates} 
Before directly comparing the different age estimates, one should bear in
mind that the dating methods may be measuring different time-scales and
may have large errors associated with them.  As mentioned above, the
stellar spectroscopy methods measure the central `luminosity weighted
average age' which means they are dominated by the last burst of star
formation, although the old stellar population also contributes. Thus the
true age of the starburst may be slightly less than the spectroscopic age.
The young GCs were possibly formed in the same star formation event as the
galaxy starburst, and as such should give a similar age as the stellar
spectroscopy. The age estimate from the \scite{FP1} trend is also based on
a star formation time-scale.

The stellar dynamics, fine structure and tail fraction ages estimate the
time since the last merger event. The `dynamical/structural' estimates are
reasonably consistent with the `starburst' ages, indicating a relatively
young age for NGC~1700, and suggesting that the merger was a gaseous one.
\scite{mihos96} have shown that in a merger between two spirals with bulges
the main starburst occurs at the time of nuclear coalescence while the
tails form $\sim0.5$ Gyr earlier. If this is the case, we might expect the
`dynamical/structural' age estimates to be slightly higher than the
`starburst' ones.  We have decided to adopt an age for NGC~1700 of
$3.0\pm1.0$ Gyr as our best estimate.  This may also correspond to the time
since the nuclei of the progenitors merged.

\subsection{Globular cluster spatial properties and specific frequency}
\subsubsection{Spatial distribution}

\begin{figure}
\begin{minipage}{8.4cm}
\centering 
\psfig{file=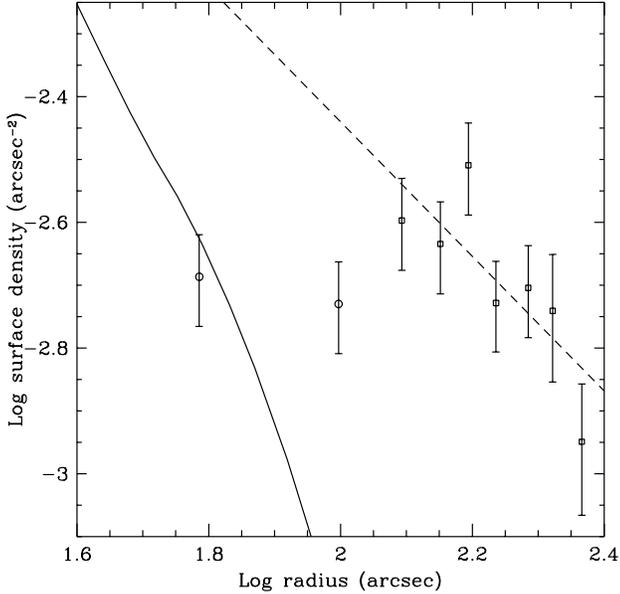,angle=0,height=8.4cm,width=8.4cm}
\end{minipage}
\caption{Surface
  density profile of our final Keck sample of globular clusters in
  NGC~1700. The significant deficit of Keck globular clusters within a
  radius of $\sim140$ arcsec (open circles) is likely to be an artefact of
  our galaxy modelling process. At larger radii, the surface density
  profile resembles a power-law (open boxes). A power-law fitted to these
  outer points is shown as the dashed line. The solid line is the stellar
  intensity profile of the underlying galaxy and has a significantly
  steeper slope than the globular cluster surface density distribution.}
\label{fig:sdprofilek}
\end{figure}

\begin{figure}
\begin{minipage}{8.4cm}
\centering
\psfig{file=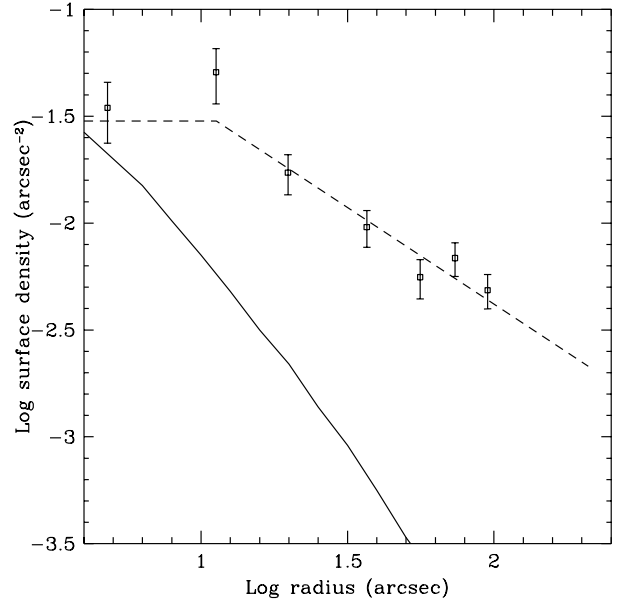,angle=0,height=8.4cm,width=8.4cm}
\end{minipage}
\caption{Surface
  density profile of our final \emph{HST} sample of globular clusters in
  NGC~1700 (open squares).  The dashed line is a power-law fitted to the data
  with a flat core region inside a radius of 11 arcsec. The solid line is
  the stellar intensity profile of the underlying galaxy and has a
  significantly steeper slope than the globular cluster surface density
  distribution.}
\label{fig:sdprofileh}
\end{figure}

Given the field--of--view, our Keck observations are ideal for defining the
outer reaches of the GC system.  We have calculated the surface density
(SD) profile for the GC system in the Keck images within 9 annuli centred
on the galaxy. For objects in the corner of the CCD, the SD was calculated
by taking into account the area of the annulus `missing' off the edges of
the chip. This method allowed us to calculate the density out to a radius
of 232 arcsec ($\sim60$ kpc). The resulting surface density profile is
shown in Fig.~\ref{fig:sdprofilek}. The error bars simply reflect the
Poisson errors on the number of GCs in each bin. At large radii the surface
density decreases like a power-law with radius (open squares). At radii
$<140$ arcsec however, there appears to be a significant deficit in the
number of GCs detected in our Keck images (shown by open circles). The most
likely cause for this is the fact that GCs at small radii are superimposed
on the bright body of the galaxy which also possesses a steep radial
gradient at these distances.  Although an elliptical model of the galaxy
was subtracted from each of the initial images, this effect seems to have
caused a reduction in our ability to detect GCs as our \emph{HST} surface density
continues to rise towards the centre at these radii as shown in
Fig.~\ref{fig:sdprofileh}. In addition, we were unable to detect GCs in the
Keck images within the central 33 arcsec due to the large saturated region
at the centre of the $I$ band image. To quantify the outer SD profile of
the GCs detected in our Keck images, we fitted the outer-most points in
Fig.~\ref{fig:sdprofilek}  with a function of the form
$\rho=\rho_{0}r^{\alpha}$. This fit is shown by the dashed line in
Fig.~\ref{fig:sdprofilek}. We find that $\rho_{0}=0.5\pm3.3$ and
$\alpha=-1.07\pm0.25$. Background contamination of the sample would add a
constant SD level to the profile. Although this contamination is likely to
be small, ideally it would be taken into consideration when computing the
fit. However, our data are not sufficient to calculate the level of
contamination and the calculated slope of the profile should be treated
with caution. The solid line in Fig.~\ref{fig:sdprofilek} represents the
stellar profile of the galaxy. The galaxy profile used is not in the usual
units of surface brightness but has been converted to
\mbox{log$(intensity)$} and arbitrarily shifted in the Y direction to allow
simple comparison of the slopes. We measure a slope of $-1.9\pm0.1$ for the
galaxy profile using the measured intensities at all radii. It thus
appears that the GC profile is flatter than the underlying stellar profile.

In order to investigate the SD profile within the central regions of
the galaxy, we use the \emph{HST} images. This profile is shown in
Fig.~\ref{fig:sdprofileh} after correction for the areas of the
annular bins missing from the WFPC2 field--of--view.  Each annulus
contains $\sim25$ GCs with the exception of the two inner bins which
contain $\sim10$ GCs. We find that the \emph{HST} surface density
follows a power-law profile exterior to $r\sim11$ arcsec (i.e. the
outer 6 data points). We again fitted these points with a function of
the form $\rho=\rho_{0}r^{\alpha}$ and find $\alpha=-0.90\pm0.15$ and
$\rho_0=0.27\pm0.19$ GC \mbox{arcsec$^{-2}$}. The slope of this fit is
within the range of values measured for other ellipticals by other
authors (e.g. \pcite{forbes97}) and also consistent with the slope
calculated from the Keck sample. Also shown in
Fig.~\ref{fig:sdprofileh} is the stellar profile of the galaxy. This
has a slope of $-1.9\pm0.1$. The GC profile is significantly flatter
than that of the galaxy, which is often the case in other systems
(e.g. Grillmair, Pritchet \& van~den~Bergh
1986; \pcite{harris86,lauer86,forbes98,ashman98}). Within $\sim11$
arcsec there is evidence for a flattening of the profile indicating
the presence of a `core region'. The core radius of the NGC~1700 GC
distribution was also measured by \scite{forbes96} to be $2.7\pm1.6$
kpc corresponding to $\sim11$ arcsec, although only 39 GCs were
detected. The presence of a core region is also commonly seen in other
ellipticals and \scite{forbes96} define a relationship between GC
system core radius and parent galaxy luminosity. This relation is in
the sense that more luminous ellipticals have more extended GC
systems. An elliptical with the luminosity of NGC~1700 (i.e.
$M_{\mathrm{V}}=-22.47$) is expected to have a core radius of $\sim3$
kpc, corresponding to $\sim12$ arcsec, although there is a large
uncertainty on this value \cite{forbes96}.  Our data are thus
consistent with the expectation from other ellipticals and with the
previous direct measurement of the NGC~1700 GC system core radius.

\subsubsection{Total number of globular clusters}
We next make an estimate of the total number, $N_{\mathrm{T}}$, of GCs
possessed by NGC~1700.  In order to estimate $N_{\mathrm{T}}$ we integrate
under the surface density profile out to some limiting radius.  We will
assume a flat profile interior to our second data point ($r=11.3$ arcsec)
with $\log({\mathrm{SD}})=-1.52$ (as shown in Fig.~\ref{fig:sdprofileh}),
although this assumption has little effect on the final calculation.
Outside this radius we assume a power-law profile.  The faint magnitude
limit for our Keck data ($V\sim25$) is much brighter than the expected peak
magnitude of a standard GC luminosity function at the distance of NGC~1700.
Thus the correction for incompleteness would be very large. In addition,
the Keck data has a higher contamination rate than \emph{HST} which is
difficult to account for without a corresponding `blank sky' image. We
therefore integrate under the power-law fitted to the \emph{HST} points
only, for which we are confident there is very little contamination, and we
expect a much smaller incompleteness correction.  We use the Keck data to
help us to define a reasonable outer limiting radius.

The dominant sources of error in determining $N_{\mathrm{T}}$ are the
choice of the outer limiting radius and the correction for
incompleteness at faint magnitudes. Fig.~\ref{fig:sdprofilek}
indicates that the surface density declines out to a radius of
$\sim210$ arcsec outside of which there is a large drop in surface
density. This suggests that the limit of the GC distribution has been
reached. We thus chose to integrate out to 210 arcsec, which
corresponds to a galactocentric distance of 50 kpc, in order to
calculate the total number of GCs within this radius. This limiting
radius is comparable to the extents of the GC systems seen in other
ellipticals (see e.g \pcite{hanes86}).  The resulting number of GCs
was then corrected to account for the lack of completeness in the
\emph{HST} sample at faint magnitudes. Based on our limiting magnitude
we estimate this correction factor to lie in the range of 2--3, thus
adopting a correction factor of 2.5, with an uncertainty of $\pm0.5$.
We estimate that we are detecting roughly $(40\pm8)$ per cent of the
total GC system within the limits of the \emph{HST} field--of--view.
We multiplied the number of GCs calculated from the SD profile by 2.5
to obtain a completeness corrected value for $N_{\mathrm{T}}$ within
50 kpc of $N_{\mathrm{T}}=1320\pm270$.  Our value is greater than that
found by
\scite{whitmore97} who estimated a total number of $528\pm48$ GCs.  However
they considered only those GCs out to the radius of the \emph{HST}
field--of--view, whereas our Keck data indicates that the GCs are still
present out to $\sim200$ arcsec.

\subsubsection{Specific frequency}  
The globular cluster specific frequency, $S_N$, is a useful quantity as it
can provide valuable constraints on the formation mechanisms of the GC
system as well as providing information on the formation history of the
parent galaxy and potentially the nature of the progenitor spirals. From
$N_{\mathrm{T}}$, we calculate the total GC specific frequency
$S_N({\mathrm{total}})$ within 50 kpc from the galaxy centre, from the
  following relation \cite{harris81}

\[
S_{N}({\mathrm{total}})=N_{\mathrm{T}}\times10^{0.4(M_{\mathrm{V}}+15)}
\]

where $M_{\mathrm{V}}$ is the absolute magnitude of the stars associated with the
population of GCs in question. In the case of $S_N({\mathrm{total}})$, the
required luminosity is the absolute magnitude of the galaxy, i.e.
$M_{\mathrm{V}}=-22.47$. We will assume a 20 per cent error in the distance to
NGC~1700, corresponding to an error in $M_{\mathrm{V}}$ of 0.4 mag. This yields for
$N_{\mathrm{T}}=1320\pm270$, a total specific frequency of
$S_N({\mathrm{total}})=1.4^{+1.0}_{-0.6}$. Elliptical galaxies have a wide
range of GC specific frequencies. The mean value for ellipticals in the
compilation of \scite{ashman98} is $S_N=5.1\pm0.6$, though there are a
couple of cases of ellipticals with $S_N$ values less than 1.0. Although
possessing large errors, our $S_N({\mathrm{total}})$ is relatively low
compared to most typical ellipticals and is more consistent with disc
galaxies.

It is clear from the age estimates discussed herein that NGC~1700 is a
relatively young galaxy. Will the GC population eventually come to resemble
those seen around typical old ellipticals ? To address this question we use
the Worthey models to predict the total specific frequency once sufficient
time has passed for NGC~1700 to have an age comparable to those of typical
elliptical galaxies today. Old, present day ellipticals have stellar
population ages of 10--15 Gyr so in order to make this comparison we
compute the luminosity of a young starburst component after it has aged by
10 Gyr. As the mass of the starburst is uncertain, we consider two cases:
Firstly we assume a starburst that represents 10 per cent of the galaxy by
mass and is embedded in a 10 Gyr population making up the remaining 90 per
cent of the galactic stellar mass.  After 10 Gyr, the models indicate that
the young stellar population will fade by $\Delta V=1.48$ mag (for an
$[{\mathit{Fe}/H}]=+0.5$ population).  As this population constitutes only 10
per cent of the galaxy mass however, the global fading is only $\Delta
V=0.18$ mag.  This results in a predicted $S_N({\mathrm{total}})$ of
$\sim1.6$, which is still lower than that expected for most typical
present day ellipticals and suggests that NGC~1700 may form a relatively
`globular cluster poor' elliptical galaxy.  If the starburst fraction is
less than $10$ per cent (as might be expected in an elliptical plus spiral
merger), then the global fading is further reduced.

Secondly, we consider a starburst that constitutes 50 per cent of the
galaxy mass. The remaining 50 per cent is made up of the old stellar
population. After 10 Gyr, the fading of the young stellar population
results in a global fading of $\Delta V=0.81$ mag. The
$S_N({\mathrm{total}})$ predicted after 10 Gyr is $\sim2.8$. This is more
consistent with elliptical galaxy specific frequencies, though still lower
than the mean $S_N$ value calculated from \scite{ashman98}.

The merger model of GC formation by \scite{ashman92} states that the
number of new globular clusters formed in a merger will be
proportional to the available gas mass, and that most `normal' old
ellipticals were formed from progenitors which were relatively
gas--rich compared to present day spirals. It is interesting to note
therefore that \emph{if} one accepts the merger hypothesis and the
fact that NGC~1700 is a relatively young elliptical, its GC specific
frequency may be expected to be relatively low. Note, however that our
`predicted' values of $S_N$ should be considered as very approximate
as there is not only an inherent difficulty in estimating our sample
completeness (and hence the number of GCs), but also a large
uncertainty associated with the assumptions made regarding the future
evolution of the galaxy luminosity.

\section{Conclusions}

We have presented results from new $B$, $V$ and $I$ band imaging of
NGC~1700, taken using the Keck telescope and reanalysed previous \emph{HST}
imaging. From the morphology of the galaxy and photometry of its globular
cluster system, we have derived new estimates of the galaxy age.

Subtraction of an elliptical model from the Keck images revealed the
presence of two symmetric tidal tail-like structures extending $\sim40$
kpc to the North-West and South-East of the galaxy.  The presence of
tidal tails is thought to be a classic signature that a galaxy has undergone a
merger event, involving the collision of two spiral galaxies during the
last few Gyr. If the observed tidal features are indeed genuine tidal
tails, then this would suggest that NGC~1700 has undergone a major merger
event during its recent history. If they are merely plumes or other fine
structure, then the situation of a disc galaxy merging with an elliptical
is also possible.  Based on the fraction of galaxy light contained within
these tails we have estimated their age at $3.2\pm1.5$ Gyr.  We have also
shown that NGC~1700 possesses boxy isophotes for radii of $7.5<r<20.0$ kpc,
and that this boxiness is largely caused by the presence of the tail
structures.

From the \emph{HST} imaging, we detected 146 globular clusters.  Of
these, 34 were in common with our GC candidates detected in the Keck
data. We then used these common GCs to refine our Keck GC selection.
After further restricting our sample in magnitude and colour we
obtained a final Keck object list of 312 GCs. These show a bimodal
colour distribution with peaks at $B-I=1.54$ and 1.98.  The colour of
the blue population is consistent with that of Galactic GCs.  Assuming
that the blue population is indeed an old metal poor system and
measuring the offset in magnitude and colour between the two
populations, we find that the red GCs are younger and more metal
rich. The \scite{bc96} stellar population models suggest that the red
population is 2.5--5.0 Gyr old and has super-solar metallicity. The
equivalent models of \scite{wortheymod} predict a significantly older
age (5.0--8.0 Gyr) and a lower metallicity
($[{\mathit{Fe}}/H]\sim-0.2$). The bimodality in $B-I$ is supported by
a hint of bimodality seen in the $V-I$ colours from \emph{HST} data.

We discuss other age estimates from the literature based on fine structure,
globular cluster colours, Fundamental Plane residuals, stellar dynamics, and
spectroscopic line strengths. We find that, although possessing significant
errors, the various age estimates generally indicate a young age for
NGC~1700 of about $3.0\pm1.0$ Gyr. The fact that the `dynamical/structural' and
`star formation' ages are similar suggests that NGC~1700 has undergone an
episode of enhanced star formation triggered by a merger event, which may
have created the galaxy itself.

The surface density profile of GCs reveals a flatter (i.e.  more extended)
profile than the underlying galaxy starlight. The total number of GCs and
present total specific frequency are estimated to be
$N_{\mathrm{T}}=1320\pm270$ and $S_N({\mathrm{total}})=1.4^{+1.0}_{-0.6}$
respectively. The value for $S_N({\mathrm{total}})$, even considering the
large errors, is quite low compared to the majority of typical ellipticals
and is more consistent with spiral galaxies. We predict that after 10 Gyr
the total $S_N$ will have increased to a maximum of $\sim2.8$ if NGC~1700
has undergone a starburst that constitutes 50 per cent of its stellar mass.
This is consistent with low $S_N$ ellipticals. If the starburst constitutes
only 10 per cent of the total stellar mass, the predicted $S_N$ is somewhat
lower at $S_N\sim1.6$. This suggests that NGC~1700 will form a relatively
`globular cluster poor' elliptical galaxy once it reaches a comparable age
to typical `old' ellipticals.

\section*{Acknowledgments}
We thank Trevor Ponman, Alejandro Terlevich and Edward Lloyd--Davies
for help and useful discussions. We also thank the referee, Keith
Ashman for his helpful comments.

Some of the data presented herein were obtained at the W.~M.~Keck
Observatory, which is operated as a scientific partnership among the
California Institute of Technology, the University of California and the
National Aeronautics and Space Administration. The Observatory was made
possible by the generous financial support of the W.~M.~Keck Foundation.

This work was also based on observations with the NASA/ESA \emph{Hubble
  Space Telescope}, obtained from the data archive at the Space Telescope
Science Institute, which is operated by the Association of Universities for
Research in Astronomy, Inc.  under NASA contract No.  NAS5-26555.

Part of this research was funded by NATO Collaborative Research grant
CRG~971552.

\nocite{rc3} \nocite{rsa} \nocite{lris}
\nocite{kmm} \nocite{bbf92} \nocite{bender94}
\nocite{couture90} \nocite{fisher96} \nocite{forbes97}
\nocite{FP1} \nocite{franx89a} \nocite{franx89b}
\nocite{geisler96} \nocite{grillmair86} \nocite{statler96}

\bibliography{ngc1700}

\begin{thebibliography}{Fritze~v. Alvensleben \& Burkert<1995>}

\bibitem[Ashman \& Zepf<1992>]{ashman92}
Ashman~K.~M., Zepf~S.~E., 1992, ApJ, 384, 50

\bibitem[Ashman \& Zepf<1998>]{ashman98}
Ashman~K.~M., Zepf~S.~E., 1998, {\em Globular Cluster Systems}.
\newblock Cambridge University Press, Cambridge

\bibitem[Ashman {\rm et~al.}<1994>]{kmm}
Ashman~K.~M., Bird~C.~M., Zepf~S.~E., 1994, AJ, 108, 2348

\bibitem[Bahcall \& Soneira<1981>]{bahcall81}
Bahcall~J.~N., Soneira~R.~M., 1981, ApJS, 47, 357

\bibitem[Balcells \& Quinn<1990>]{balcells90}
Balcells~M., Quinn~P.~J., 1990, ApJ, 361, 381

\bibitem[Bender {\rm et~al.}<1992>]{bbf92}
Bender~R., Burstein~D., Faber~S.~M., 1992, ApJ, 399, 462

\bibitem[Bender {\rm et~al.}<1994>]{bender94}
Bender~R., Saglia~R.~P., Gerhard~O.~E., 1994, MNRAS, 269, 785

\bibitem[Bertin \& Arnouts<1996>]{bertin96}
Bertin~E., Arnouts~S., 1996, A\&AS, 117, 393

\bibitem[Bruzual \& Charlot<2000>]{bc96}
Bruzual~G.~A., Charlot~S.
\newblock 2000.
\newblock In preparation

\bibitem[Couture {\rm et~al.}<1990>]{couture90}
Couture~J., Harris~W.~E., Allwright~J. W.~B., 1990, ApJS, 73, 671

\bibitem[de~Vaucouleurs {\rm et~al.}<1991>]{rc3}
de~Vaucouleurs~G., de~Vaucouleurs~A., Corwin~Jr.~H.~G., Buta~R.~J., Paturel~G.,
  Fouque~P., 1991, {\em Third Reference Catalogue of Bright Galaxies}.
\newblock Springer, Berlin, (RC3)

\bibitem[Fisher {\rm et~al.}<1996>]{fisher96}
Fisher~D., Franx~M., Illingworth~G., 1996, ApJ, 459, 110

\bibitem[Forbes \& Thomson<1992>]{ft92}
Forbes~D.~A., Thomson~R.~C., 1992, MNRAS, 254, 723

\bibitem[Forbes {\rm et~al.}<1996>]{forbes96}
Forbes~D.~A., Franx~M., Illingworth~G.~D., Carollo~C.~M., 1996, ApJ, 467, 126

\bibitem[Forbes {\rm et~al.}<1997>]{forbes97}
Forbes~D.~A., Brodie~J.~P., Grillmair~C.~J., 1997, AJ, 113, 1652

\bibitem[Forbes {\rm et~al.}<1998a>]{forbes98}
Forbes~D.~A., Grillmair~C.~J., Williger~G.~M., Elson~R. A.~W., Brodie~J.~P.,
  1998a, MNRAS, 293, 325

\bibitem[Forbes {\rm et~al.}<1998b>]{FP1}
Forbes~D.~A., Ponman~T.~J., Brown~R. J.~N., 1998b, ApJ, 508, L43

\bibitem[Franx {\rm et~al.}<1989a>]{franx89a}
Franx~M., Illingworth~G., Heckman~T., 1989a, ApJ, 344, 613

\bibitem[Franx {\rm et~al.}<1989b>]{franx89b}
Franx~M., Illingworth~G., Heckman~T., 1989b, AJ, 98, 538

\bibitem[Fritze~v. Alvensleben \& Burkert<1995>]{fritze95}
Fritze~v. Alvensleben~U., Burkert~A., 1995, A\&A, 300, 58

\bibitem[Gebhardt \& Kissler-Patig<1999>]{gebhardt99}
Gebhardt~K., Kissler-Patig~M., 1999, AJ, 118, 1526

\bibitem[Geisler {\rm et~al.}<1996>]{geisler96}
Geisler~D., Lee~M.~G., Kim~E., 1996, AJ, 111, 1529

\bibitem[Gonzalez<1992>]{gonzalez92}
Gonzalez~J., 1992, {\rm PhD thesis}, University of California

\bibitem[Goudfrooij {\rm et~al.}<1994>]{g94}
Goudfrooij~P., Hansen~L., Jorgensen~H.~E., Norgaard-Nielson~H.~U., de~Jong~T.,
  van~den Hoek~L.~B., 1994, A\&AS, 104, 179

\bibitem[Grillmair {\rm et~al.}<1986>]{grillmair86}
Grillmair~C., Pritchet~C., van~den Bergh~S., 1986, AJ, 91, 1328

\bibitem[Hanes \& Harris<1986>]{hanes86}
Hanes~D.~A., Harris~W.~E., 1986, ApJ, 309, 564

\bibitem[Harris \& van~den Bergh<1981>]{harris81}
Harris~W.~E., van~den Bergh~S., 1981, AJ, 86, 1627

\bibitem[Harris<1986>]{harris86}
Harris~W.~E., 1986, AJ, 91, 822

\bibitem[Hartigan \& Hartigan<1985>]{hartigan85}
Hartigan~J.~A., Hartigan~P.~M., 1985, Ann. Stat., 13, 70

\bibitem[Holtzman {\rm et~al.}<1995>]{holtzman95}
Holtzman~J.~A., Burrows~C.~J., Casertano~S., Hester~J.~J., Trauger~J.~T.,
  Watson~A.~M., Worthey~G., 1995, PASP, 107, 1065

\bibitem[Keel \& Wu<1995>]{keelwu95}
Keel~W.~C., Wu~W., 1995, AJ, 110, 129

\bibitem[Koo \& Kron<1992>]{kookron92}
Koo~D.~C., Kron~R.~G., 1992, ARA\&A, 30, 613

\bibitem[Landolt<1992>]{landolt92}
Landolt~A.~U., 1992, AJ, 104, 340

\bibitem[Lauer \& Kormendy<1986>]{lauer86}
Lauer~T.~R., Kormendy~J., 1986, ApJ, 303, L1

\bibitem[Mihos \& Hernquist<1996>]{mihos96}
Mihos~C.~J., Hernquist~L., 1996, ApJ, 464, 641

\bibitem[Oke {\rm et~al.}<1995>]{lris}
Oke~J.~B. {\rm et~al.}, 1995, PASP, 107, 375

\bibitem[Poulain<1988>]{poulain88}
Poulain~P., 1988, A\&AS, 72, 215

\bibitem[Prugniel \& Simien<1996>]{prugniel96}
Prugniel~P., Simien~F., 1996, A\&A, 309, 749

\bibitem[Sandage \& Tammann<1981>]{rsa}
Sandage~A., Tammann~G.~A., 1981, {\em A Revised Shapley-Ames Catalog of Bright
  Galaxies}.
\newblock Carnegie Institution, Washington, (RSA)

\bibitem[Schweizer \& Seitzer<1992>]{ss92}
Schweizer~F., Seitzer~P., 1992, AJ, 104, 1039

\bibitem[Statler {\rm et~al.}<1996>]{statler96}
Statler~T.~S., Smecker-Hane~T., Cecil~G.~N., 1996, AJ, 111, 1512

\bibitem[Thomson \& Wright<1990>]{thomson90}
Thomson~R.~C., Wright~A.~E., 1990, MNRAS, 247, 122

\bibitem[Toomre \& Toomre<1972>]{toomre72}
Toomre~A., Toomre~J., 1972, ApJ, 178, 623

\bibitem[Whitmore {\rm et~al.}<1997>]{whitmore97}
Whitmore~B.~C., Miller~B.~W., Schweizer~F., Fall~S.~M., 1997, AJ, 114, 1797

\bibitem[Worthey<1994>]{wortheymod}
Worthey~G., 1994, ApJS, 95, 107

\bibitem[Yoshii<1993>]{yoshii93}
Yoshii~Y., 1993, ApJ, 403, 552

\bibitem[Zepf \& Ashman<1993>]{zepf93}
Zepf~S.~E., Ashman~K.~M., 1993, MNRAS, 264, 611

\end{thebibliography}

\label{lastpage}

\end{document}